# Defect-induced nonlinearity in 2D nanoparticles

JIE XU[1] AND ERIC PLUM[1],*

1. Optoelectronics Research Centre and Centre for Photonic Metamaterials, University of Southampton, Southampton, SO17 1BJ, United Kingdom
*erp@orc.soton.ac.uk

**Abstract:** Optical nonlinearity depends on symmetry and symmetries vanish in the presence of defects. Vaccancy defects in centrosymmetric crystals and thin films are a well-known source of even-order optical nonlinearity, e.g. causing second harmonic generation. The emerging ability to manipulate defects in two-dimensional materials and nanoparticles provides an opportunity for engineering of optical nonlinearity. Here, we demonstrate the effect of defects on the nonlinear optical response of two-dimensional dielectric nanoparticles. Using a toy model, where bound optical electrons of linear atoms are coupled by nonlinear Coulomb interactions, we model defect-induced nonlinearity. We find that defects at particle edges contribute strongly to even-order optical nonlinearity and that unique nonlinear signatures of different defect states could provide the smallest conceivable QR-codes and extremely high density optical data storage, in principle approaching 1 bit per atom.

## 1. Introduction

Harmonic generation is of key importance for wavelength conversion of optical signals and optical information processing. In the electric dipole approximation, media with inversion centre cannot generate even order harmonics. However, defects can break the inversion centre and enable even order harmonic generation in centrosymmetric media. Harmonic generation as well as wave mixing in nonlinear photonic crystals with defects have attracted attention [1-4] and defect-enhanced second harmonic generation (SHG) has been observed in various materials, including crystals [5, 6], 2D materials [7], doped fibres [8], semiconductors [9], their superlattices [10, 11] and interfaces [12, 13]. Recent advances in micromachining [14, 15], nanoparticle decoration [16, 17], atomic healing [18], lateral heterostructures [19, 20] and heterocrystals [21] have established the feasibility of defect manipulation and postprocessing, which provides an opportunity for the development of two-dimensional (2D) materials with engineered optical nonlinearity [22]. Second harmonic generation (SHG) has been used for monitoring of defects engineering in transition metal dichalcogenides [23] and controlling nonlinear properties through defect engineering has been reported [24, 25]. Effects of vacancy defect concentration on high-order harmonic generation have also been investigated [26-28].

In this paper, we present a model of defect-induced harmonic generation in 2D dielectric nanoparticles described as a lattice of linear oscillators coupled by Coulomb interactions. The model shows how specific defects affect the particle's even and odd order optical nonlinearities (Fig. 1). Our results indicate that defects at particle edges have the largest influence on optical nonlinearity. Different nonlinear optical responses of different defect arrangements indicate that information could – in principle – be encoded in atomic defects and read optically, via their harmonic generation signature. This would allow nanoparticles with defects to serve as the smallest conceivable QR-codes. The fundamental information density limit of such optical data storage would not be determined by diffraction, but rather by the spacing of atoms. We illustrate this by showing that harmonic generation is controlled by the location of defects in a nanoparticle and by demonstrating harmonic-based recognition of nano-QR-codes representing different characters. Such information could be written by AFM-based techniques and read by scanning near-field microscopy techniques.

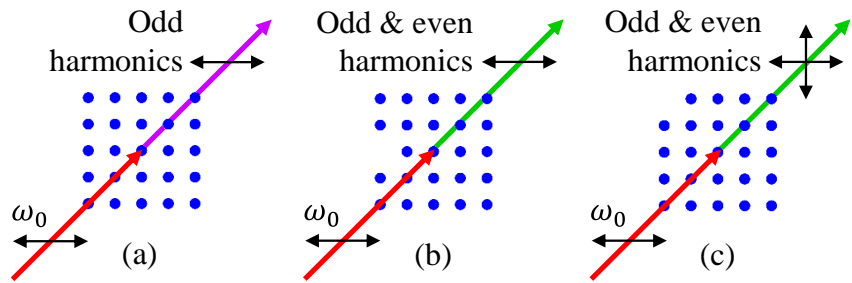

**Fig. 1.** Defect-induced optical nonlinearity. (a) Along the optical axis, as shown, a square particle will only generate odd harmonics without polarization change. The presence of defects enables even harmonic generation (b) without and (c) with polarization change.

## 2. Results and discussion

*2.1. Modelling defect-induced optical nonlinearity of 2D nanoparticles*

In the confined geometry of a two-dimensional nanoparticle, the collective nonlinear response of the atomic array can arise from the Coulomb interactions [29]. We describe atoms as classical Lorentz oscillators [30], consisting of one optical electron bound to one nucleus. For an individual atom, the interaction between the electron with coordinate $r$(t) and its stationary nucleus at $R$ is described by a linear restoring force. The electromagnetic response of the single atom is strictly linear and arises from a harmonic potential [29]. However, when an electric field is applied to a particle of $N$ atoms, the optical electrons start to oscillate and these oscillations are affected by Coulomb interactions with other electrons and other cores, introducing a nonlinear term in the equation of motion for each atom:

$$\ddot{r}_k + \gamma \dot{r}_k + \omega_0^2 \left( r_k - R_k \right) = \frac{q}{m} E(t) + \frac{q^2}{4\pi\varepsilon_0 m} \sum_{i \neq k}^{N} \left( \frac{r_k - r_i}{|r_k - r_i|^3} - \frac{r_k - R_i}{|r_k - R_i|^3} \right) \quad (1)$$

where $r_k$ is the displacement of electron $k$ in the particle; $\gamma$ is the damping frequency; $q$ and $m$ are the electron charge and mass; and $\omega_0$ is the angular resonance frequency of an isolated atomic harmonic oscillator. The resonance angular frequency of a single atom $\omega_0$ is set to $9.4\times10^{15}$ rad/s, corresponding to a resonance at 200 nm wavelength in the UV band, and the damping frequency is chosen as $\gamma = 0.01\omega_0$ to approximate a typical atomic response in dielectrics (such as ITO, TiOx and SiN). The spacing between atoms is chosen to be 0.5 nm. The pumping electric field angular frequency is $0.19\omega_0$, i.e. far from the atomic resonance, and corresponds to the 1064 nm wavelength of an Nd:YAG laser. The pumping electric field magnitude ($E_0$) is set to $8.68\times10^{10}$ V/m to explore physical trends and higher-order nonlinear optical effects in a regime, where they are sufficiently large to make numerical errors of differential equation solvers irrelevant.

We consider the radiation generated along $z$ by 2D structures in the *xy*-plane (nanoparticle plane) in response to pumping along $z$, to mimic the structure's nonlinear optical response along the propagation direction for pumping at normal incidence. In this case, radiation generated by the particle is determined only by its total (net) electric dipole moment. Higher multiples cannot contribute to the radiation along the $z$ direction. The displacement of each optical electron in the particle is calculated in the time domain by solving their coupled equations of motion [Equation (1)]. We consider a square particle cut from a square lattice (main text) and a hexagonal particle cut from a hexagonal lattice (Supplement 1), introducing defects by removing atoms at specific positions in the lattice. The coupled system of differential equations is solved in Matlab starting without optical electron displacement and allowing transitional effects to pass before analysing the oscillation of all electrons over 1000 periods of the driving field. The electric dipole moment of each atom $d_k$ is the charge times the electron displacement. The sum of these dipoles over all atoms, $P=\sum_k d_k$, gives the total electric dipole moment $P(t)$ of the nanostructure in the time domain. The total electric dipole moment is separated into linear and nonlinear components, $P^{(1)}, P^{(2)}, P^{(3)} \ldots$, oscillating at the driving frequency $\omega_p$ and its harmonics $2\omega_p, 3\omega_p \ldots$, through Fourier transformation. In the presented Fourier series, the peak values are the dipole amplitudes in Coulomb meters at harmonic frequencies.

*2.2. Defect-induced nonlinearity in a square nanoparticle*

Since radiation of even order harmonics along $z$ is forbidden in structures with even-fold rotational symmetry [31, 32], one can easily identify the defect-induced nonlinearity of such structures at even harmonics. Therefore, we first study a square particle with 25 atoms (D4 symmetry, with square lattice) and then introduce one vacancy defect at three different positions. Figure 2a illustrates harmonic generation in the square particle. The frequency-dependence of the particle's total electric dipole moment $P$ generated by electric pump field $E_0\cos(\omega_p t)$ is shown, where $P_{\alpha,\beta}$ refers to the $\alpha$-component of the dipole moment caused by a $\beta$-polarized pump field and $\alpha,\beta$ is $x$ or $y$. As required by symmetry, $x$-polarized excitation generates a dipole only along $x$, while $y$-polarized excitation generates a dipole only along $y$ with the same magnitude. i.e. $P_{x,x}=P_{y,y}$, while $P_{x,y}$ and $P_{y,x}$ are forbidden. As expected for a particle with even-fold rotational symmetry [31, 32], we observe only odd harmonics – $P^{(1)}, P^{(3)}, P^{(5)}$ – which appear as peaks in Fig. 2a. However, when there is a vacancy defect in the particle, even harmonics can occur, as shown by Fig. 2(b-d).

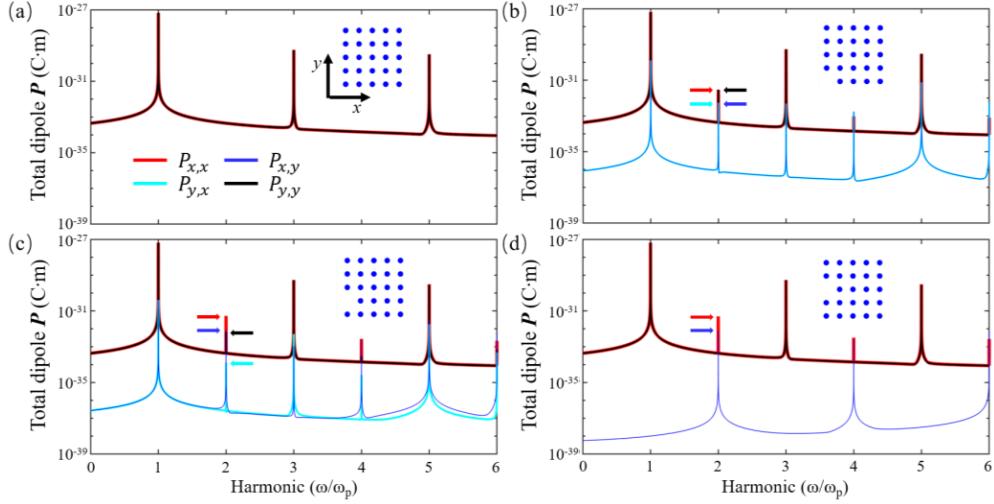

**Fig. 2.** Harmonic generation in a square nanoparticle without and with defects. (a)-(d) Frequency dependence of the electric dipole moment of a square particle (insert) with no defect (a), and (b-d) one defect at three different locations, in response to optical pumping at frequency $\omega_p$. $P_{\alpha,\beta}$ indicates the $\alpha$-component of the particle's electric dipole moment caused by $\beta$-polarized pumping. Arrows indicate the magnitudes of $P_{x,x}$, $P_{y,y}$, $P_{x,y}$ and $P_{y,x}$ at second harmonic frequency.

The defect position has a great influence on the particle's nonlinear response. When the defect occurs at a corner (Fig. 2b), $x$ or $y$-polarized excitation generates dipole moments along both the $x$ and $y$ directions, and we observe $P_{x,x}=P_{y,y}$ and $P_{x,y}=P_{y,x}$. When the defect occurs in between a corner and the middle of an edge (Fig. 2c), the four components $P_{x,x}$, $P_{y,y}$, $P_{x,y}$ and $P_{y,x}$ are all allowed and different. For a defect in the middle of an edge (Fig. 2d), only $P_{x,x}$, $P_{y,y}$ and one of $P_{x,y}$ and $P_{y,x}$ are allowed.

Vacancy defects in a hexagonal particle with 19 atoms (D6 symmetry, with hexagon lattice) have also been studied and lead to even-order optical nonlinearity in a similar way, see Fig. S1 in Supplement 1.

## 2.3. The influence of defect position on harmonic generation

The importance of the defect position was revealed by Fig. 2. Here we investigate the dependence of optical nonlinearity on the defect position systematically. Fig. 3 shows the calculated components of the total electric dipole moment $P$ as a function of the position of a single defect. The top row shows the total electric dipole moment along the $x$-direction for $x$-polarized pumping, $P_{x,x}$ at odd (left) and even (right) harmonic frequencies. At odd harmonic frequencies, $P_{x,x}$ is of almost the same magnitude for all the 25 defect positions. In contrast, $P_{x,x}$ at even harmonic frequencies depends strongly on the defect's location, being strongest for defects on the left and right edges and vanishing only for defects that do not break the $y$ symmetry axis of the nanoparticle. The bottom row shows the total dipole moment along the $y$-direction for $x$-polarized pumping, $P_{y,x}$ at odd (left) and even (right) harmonic frequencies. At odd harmonics, defects at corners induce the strongest dipole moment $P_{y,x}$, while this component does not arise from defects that maintain the horizontal or vertical mirror symmetry of the nanoparticle. In contrast, for even harmonics the total electric dipole $P_{y,x}$ is strongest for defects at top and bottom edges and vanishes only for defects that do not break the nanoparticle's $x$ symmetry axis. Similar behaviour is seen for hexagonal particles (see Fig. S2 in Supplement 1).

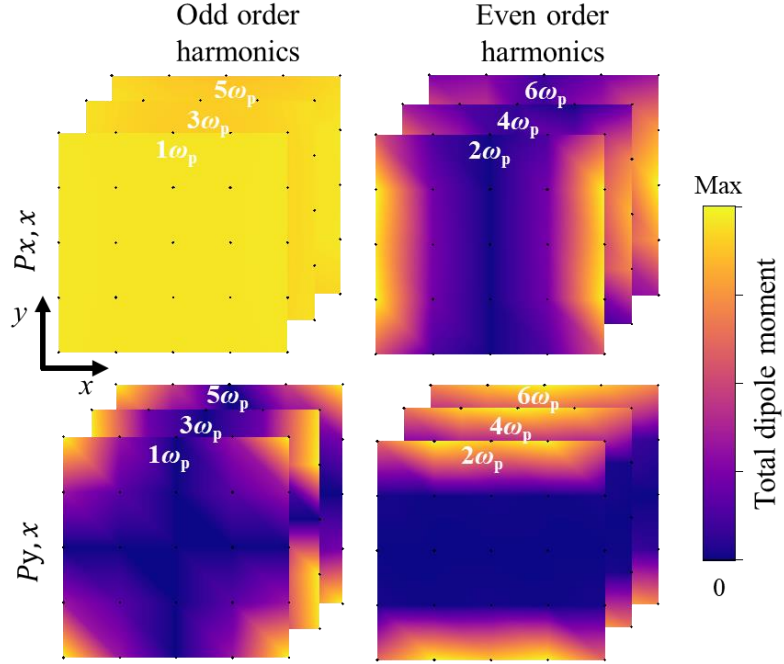

Figure 3. Total electric dipole moment **P** of a square nanoparticle as a function of the position of a single vacancy defect. The amplitude at each atom position (black dot) represents the total dipole moment of the particle when the atom is vacant. The top (bottom) row shows the dipole component parallel (orthogonal) to the *x*-polarized pump. Stacked images for different – either even or odd – harmonics show the same qualitative behaviour.

## 2.4. Harmonic generation signatures of atomic defect patterns

Recent progress in 2D material fabrication demonstrates the feasibility of defect manipulation and postprocessing techniques and SHG has already been used for monitoring defect engineering in transition metal dichalcogenides [23]. Our model can instruct defect engineering. We argue that 2D defect patterns could act as the smallest conceivable 2D barcodes, i.e. nano-QR-codes, that could be read based on their nonlinear optical properties. Fig. 4 shows the frequency dependence of the total electric dipole moment of four defect patterns – "1, 2, 3, 4" – in a square lattice in response to a pump field $E_0\cos(\omega_p t)$. We can identify each of them by looking into their second harmonic generation. Defect pattern "1" does not generate an overall electric dipole moment at the second harmonic frequency (Fig. 4a), while for pattern "3" (Fig. 4c) only the electric dipoles along *x* can be generated at the second harmonic frequency. Although all four electric dipole components are allowed for defect patterns "2" and "4" (Fig. 4b,d), components of equal amplitudes $P_{x,x}=P_{y,y}$ and $P_{x,y}=P_{y,x}$ are only observed for pattern "4". Thus, the relative strength of the second harmonic total electric dipole components can be used to identify different defect patterns. We note that other polarizations and harmonics could also be considered.

In principle, a square lattice containing *N* atom positions can have $2^N$ different defect states. Mirror symmetry with respect to *x* and *y* in Fig. 3 implies that up to four different defect states – that are related by reflections with respect to *x*, *y*, or both – share the same nonlinear signature. (Fewer in case of defect states that have one or more relevant mirror symmetries.) Therefore, $>2^{N-2}$ defect states may have distinguishable nonlinear signatures, suggesting that nonlinear detection of defect states could enable extremely high density optical data storage, in principle approaching 1 bit per atom.

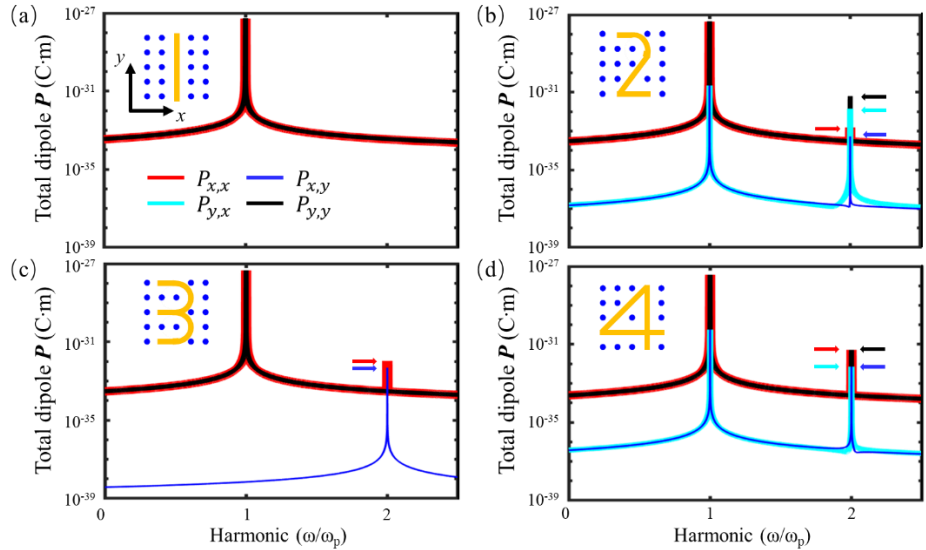

**Fig. 4.** Nano-QR-codes. Harmonic generation in a square nanoparticle with different atomic defect patterns. (a)-(d) Frequency dependence of the electric dipole moment of a square particle (insert) with defect patterns (a) "1", (b) "2", (c) "3", (d) "4" in response to optical pumping at frequency ω$_p$. $P_{α,β}$ indicates the α-component of the particle's electric dipole moment caused by β-polarized pumping.

## 3. Conclusion

In summary, we have shown that defects have a large influence on even harmonic generation by 2D nanostructures. Information could be encoded in atomic defects via defect engineering and read by its harmonic generation signature. Such information could be written by AFM-based techniques and read by scanning near-field microscopy techniques. We illustrate this by showing that harmonic generation is controlled by the location of defects in square and hexagonal particles and by demonstrating the harmonic-based recognition of nano-QR-codes representing different characters.

## Funding

This work is supported by the UK's Engineering and Physical Sciences Research Council (grants EP/M009122/1 and EP/T02643X/1), and the China Scholarship Council (CSC No. 201706310145).

## Acknowledgements

The authors thank Nikolay I. Zheludev for helpful advice.

## Disclosures

The authors declare no conflicts of interest.

## Data availability

Data from this paper is available from the University of Southampton ePrints research repository: https://doi.org/10.5258/SOTON/D1960

## Supplemental document

See Supplement 1 for supporting content.

# Supplementary information

## S1. Defect-induced nonlinearity in a hexagonal nanoparticle

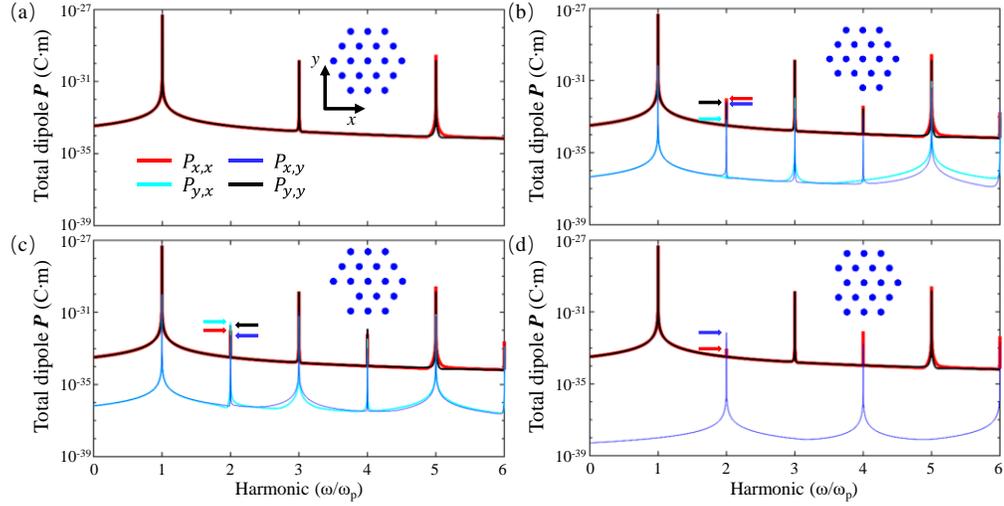

**Fig. S1.** Harmonic generation in a hexagonal nanoparticle (a) without and (b-d) with single defects in different positions of the hexagonal atomic lattice of D6 symmetry (insets). Frequency dependence of the electric dipole moment of the particles in response to optical pumping at frequency $\omega_p$. $P_{\alpha,\beta}$ indicates the $\alpha$-component of the particle's electric dipole moment caused by $\beta$-polarized pumping at normal incidence. The arrows mark the magnitudes of $P_{x,x}$, $P_{y,y}$, $P_{x,y}$ and $P_{y,x}$ at the second harmonic frequency.

The influence of vacancy defects on harmonic generation by a hexagonal nanoparticle with 19 atoms (D6 symmetry, hexagonal lattice) is shown by Fig. S1. Electric dipole moments at even order harmonics are forbidden in the hexagonal particle without defects (Fig. S1a), but emerge when a single vacancy defects breaks the symmetry of the particle (Fig. S1b-d). The amplitudes of the electric dipole components excited by $x$ and $y$ polarized pumping depend on the defect position.

## S2. Influence of defect position on harmonic generation in a hexagonal nanoparticle

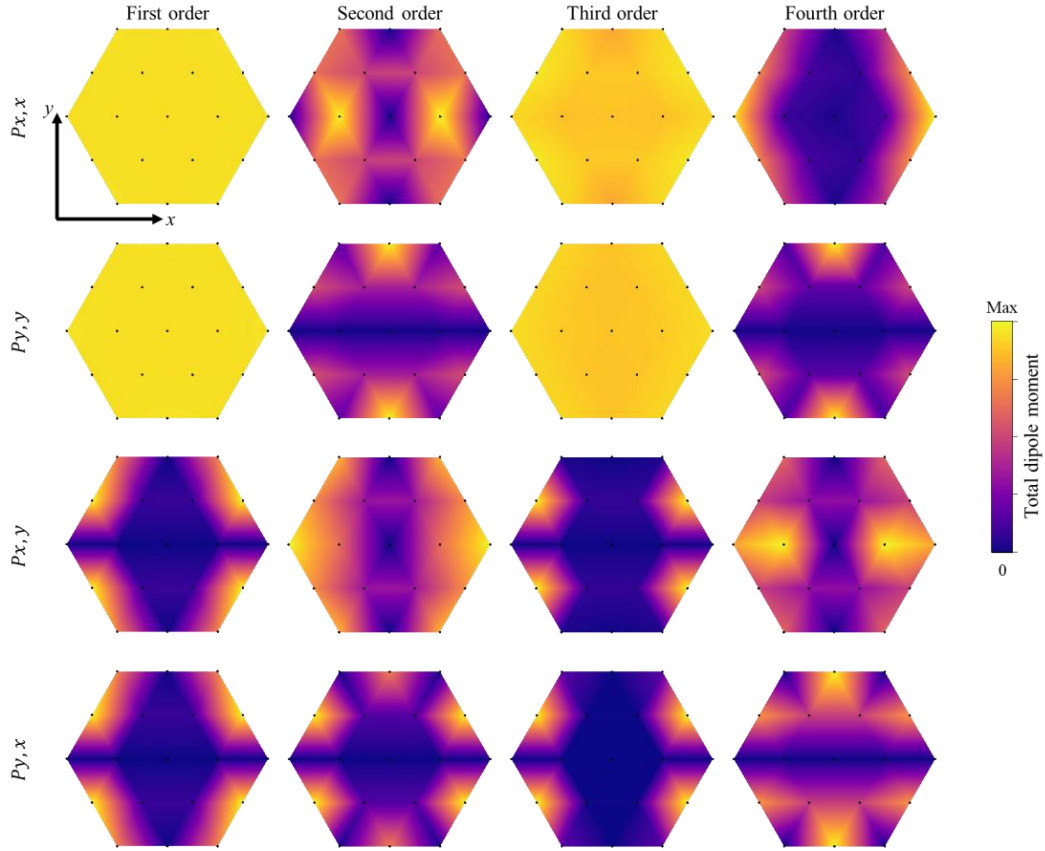

**Fig. S2.** Total electric dipole moment *P* of a hexagonal particle as a function of defect position. The amplitude at each atom (black dot) represents the total dipole moment when that atom is vacant. The four components of the particle's total electric dipole (rows) are shown for the first to fourth order harmonics (columns).

This dependence on the defect position is shown by Fig. S2. Owing to the complete particle's symmetry, that includes mirror symmetry along *x* and *y*, the distributions of total electric dipole moment as a function of defect position are mirror-symmetric along *x* and *y*. Electric dipole moments at odd harmonics, $P_{x,x}^{(1)}$, $P_{y,y}^{(1)}$, $P_{x,x}^{(3)}$ $P_{y,y}^{(3)}$, …, which can be excited without defects (Fig. S1a), do not depend strongly on the defect position. All other electric dipole components are forbidden in the complete particle and therefore depend strongly on the presence of defects in locations that affect the symmetry of the nanoparticle. Like for the square particle, pumping with *x* and *y* polarizations yields similar distributions of *odd-order* total electric dipole moment as a function of vacancy defect position. In contrast, due to lack of four-fold rotational symmetry of the hexagonal particle and its lattice, pumping with *x* and *y* polarizations yields distinctively different distributions of *even-order* total electric dipole moment as a function of defect position.